\magnification = 1200
\baselineskip = 12 pt
\input psfig.sty
\noindent
\centerline{\bf LI-RICH AGB/RGB STARS:}
\centerline{\bf LITHIUM ABUNDANCES AND MASS LOSS$^*$}
\vfootnote*{\sevenrm The 19th Cambridge 
Workshop on Cool Stars, Stellar Systems, and the Sun, ed. G. A. Feiden, 2016}
\par\noindent
\centerline{\bf W. J. Maciel, R. D. D. Costa} 
\par\noindent
\centerline{\bf University of S\~ao Paulo} 


\bigskip
\centerline{\bf Abstract}
\medskip
\vbox{\hsize = 15 true cm\baselineskip = 8 pt
\sevenrm \hangindent = 30 pt \parindent = 30 pt
Most metal-rich AGB/RGB stars present strong Li underabundances, since this element 
is easily destroyed in the high temperatures of the stellar interiors. In spite of 
this fact, several of these stars are Li-rich, having Li abundances given by 
$\epsilon$(Li) = log (Li/H)  + 12 $>$ 1.5. In the present work, we extend our previous 
investigation on AGB/RGB stars to the expected mass loss rates of these stars. 
Specifically, we look for correlations between the Li abundances 
and the mass loss rates or related parameters in Li-rich AGB/RGB stars. We have estimated 
the mass loss rates using a modified form of the Reimers formula and  applied it to a 
large sample of 104 Li-rich giant stars for which reliable stellar data are available.  
Our proposed method assumes a linear relation between the stellar luminosity and the Li 
abundance, so that the luminosity can be estimated from the Li abundance. The stellar 
mass is then obtained from the effective temperature and luminosity using recent 
evolutionary tracks. The stellar radius can be determined from the stellar gravity, so 
that the mass loss rate can be calculated using an adequate calibration  involving both 
Li-rich and Li-poor stars in the AGB/RGB branches. The results show that most Li-rich 
stars have lower mass loss rates compared with C-rich or O-rich giants that do not present 
Li enhancements.}
\bigskip
\noindent{\bf 1. Introduction}
\medskip
Most metal-rich AGB/RGB stars present strong Li underabundances, since this element is easily 
destroyed in the high temperatures of the stellar interiors. However, several of these stars 
are Li-rich, with Li abundances by number of atoms given by $\epsilon$(Li) = 
$\log$ (Li/H) + 12 $>$ 1.5. Li-rich giant stars comprise low mass stars in the Red Giant Branch 
(RGB), especially those located at the luminosity bump, and intermediate mass stars in the 
Asymptotic Giant Branch (AGB), especially those near the giant branch clump. The main production 
mechanism for $^7$Li is believed to be the Cameron-Fowler mechanism (Cameron and Fowler 
1971), in  which there is an enrichment of $^3$He following the first dredge-up episode. 
The excess $^3$He may reach the inner stellar layers, where it is burned to form $^7$Be, which 
leads to the production of $^7$Li. This element is then raised to the stellar outer layers, which 
characterizes the Li-rich phase, before being burned out or diluted into the stellar atmosphere. 

In a previous work we have considered a large sample of Li-rich stars, and investigated the Li 
abundance trends with several stellar parameters, such as metallicity, the effective temperature, 
mass, radius, and luminosity (Maciel \& Costa 2012, 2015).
In the present work, we apply the correlations involving the stellar radius and luminosity
with the Li abundance and look for correlations between the abundances and the mass loss rates 
or related parameters in Li-rich AGB/RGB stars. Comparing these results with data from Li-poor 
stars for which luminosities and mass loss rates are independently known, we derive the mass loss 
rates of Li-rich RGB/AGB stars. Section 2 discusses in detail the Li-rich and L-poor 
stellar samples considered, section 3 describes the adopted method, and section 4 presents the  
results obtained so far.
\vfill\eject

\noindent{\bf 2. The Sample}
\medskip
\noindent
2.1 Li-rich Stars
\medskip

The initial sample of Li-rich AGB/RGB stars is the one previously considered by Maciel \& Costa 
(2012, 2015), with seven additional stars from Brown et al. (1989).
As discussed in more detail by Maciel \& Costa (2012, 2015), the original 
sources of the data are Brown et al. (1989), Mallik (1999), Gonzalez et al. (2009),
Monaco et al. (2011), and Kumar et al. (2011). 
This sample was extended by the inclusion of several Li-rich stars recently published, as given by 
Lebzelter et al. (2012), Monaco et al. (2014), K\"ov\'ari et al. (2013), Martell \& Shetrone 
(2013), and Lyubimkov et al. (2012).  

Lebzelter et al. (2012) derived spectroscopic Li abundances in a sample of bulge stars 
from the bottom to the tip of the red giant branch. It is known that a few giants on the ascending RGB 
are Li-rich, as well as some stars at the bump of the luminosity function on the RGB for low mass stars 
and on the AGB for intermediate mass stars. The authors have used FLAMES spectra along with COMARCS 
atmospheres and bulge giant isochrones and derived effective temperatures, gravities and Li abundances 
from the 670.8 nm Li line. From an original sample of 401 stars, Lebzelter et al. (2012)
have found detectable Li 670.8 nm line in 31 stars, and 3 of them can be considered as Li-rich using 
the criterium mentioned earlier. These stars are located on the upper RGB, above the luminosity bump.

Monaco et al. (2014) identified a super Li-rich star in the open cluster Trumpler~5 
based on FLAMES/VLT spectra and 3D-NLTE models. The star is \# 3416, which is a core He-burning red 
clump star with $\epsilon$(Li) = 3.75. The star is located at $\alpha$(2000) = 06:36:40.2 and 
$\delta$(2000) = 09:29:47.8. 

K\"ov\'ari et al. (2013) presented Doppler imaging of two Li-rich K giants and used 
optical spectroscopy and photometry to derive their fundamental properties, such as effective 
temperature, luminosities, masses, etc. From their analysis it is concluded that both stars are 
located at the end of the first Li dredge-up on the RGB. 

Martell \& Shetrone (2013) presented a sample of 23 Li-rich field giants from the 
Sloan Digital Sky Survey with high-resolution follow up spectroscopy. These objects are located in 
the upper right region of the HR diagram, including the RGB, the AGB and the red clump. For these 
objects, high resolution spectroscopy leads to the determination of the stellar parameters. The 
stellar masses are probably in the range 1--3 M$_\odot$, where the lower limit corresponds to stars 
in the RGB, particularly those near the luminosity bump, and the upper limit corresponds to AGB 
stars (cf. Charbonnel \& Balachandran 2000).

Lyubimkov et al. (2012) derived accurate Li abundances in a sample of F, G giants 
and supergiants, out of which there are 15 objects having $\epsilon$(Li) $>$ 1.5, that is, are 
Li-rich according to the criterium adopted here. 

As we will see in the next section, in order to apply our method, the effective temperature $T_{eff}$ , 
gravity $\log g$, and Li abundance $\epsilon$(Li) of the Li-rich stars must be known. Applying this 
condition and removing some objects that lie outside the range of the adopted parameters, namely 
the Li-abundance, luminosity and effective temperature, a final sample of 104 Li-rich stars is 
obtained For details on these stars, the reader is referred to the original papers.
\bigskip
\noindent
2.2 Li-poor Stars
\medskip

There are reliable determinations of the luminosities and mass loss rates for many RGB/AGB stars 
without Li enhancements in the literature, which can be used in order to estimate the corresponding 
quantities of Li-rich stars. The Li-poor stars used as a comparison with the Li-rich objects come 
from the large samples by Gullieuszik et al. (2012) and Groenewegen et al. 
(2009).

Gullieuszik et al. (2012) obtained accurate mass loss rates and luminosities for a 
large sample of AGB stars in the LMC from the VISTA survey (VMC). Dust radiation transfer models were 
compared with the obtained spectral energy distributions (SED) from VMC data and available photometry 
from the optical to mid-infrared wavelengths. The AGB sample includes 373 objects. Excluding the 
objects for which the complete calculation could not be done by lack of data (mass loss rate 
$dM/dt$ and/or luminosity $\log L/L_\odot$), we have a final sample of 178 objects. 

Groenewegen et al. (2009) used dust radiative transfer models for a large sample 
of C-rich and O-rich AGB stars in the SMC and LMC with Spitzer data, and derived mass loss rates and 
luminosities for these objects. The O-rich stars were classified as foreground objects (FG), red 
supergiants (RSG) and AGB stars. The obtained relations involving the mass loss rates, luminosities 
and pulsation periods were further compared with predictions of models by Vassiliadis \& Wood 
(1993) as well as models based on the Reimers mass loss relation. The total C-rich 
AGB star sample included 101 objects, but the object \# 069 wbp17 has a mass loss rate several orders 
of magnitude below the other objects, so that it will be excluded, remaining 100 stars in the C-rich 
sample. The O-rich stars include 86 objects, but 10 stars are considered as foreground objects (FG) 
and about 42 objects considered as red supergiants (RSG), essentially based on the correlation between 
the bolometric magnitude and period derived by Wood et al. (1983). Excluding the FG and 
RSG objects we have 34 AGB stars in the O-rich sample. The Magellanic Clouds have the obvious advantage 
of a known distance, although their lower metallicity relative to the Milky Way may introduce some 
uncertainties when these relationships are applied to galactic objects. However, as discussed in 
Groenewegen et al. (2007), there is no clear evidence of a metallicity 
dependence of the mass loss rate for C-rich stars. The estimated uncertainties are generally of 10\% 
for the luminosity and 25\% in the mass loss rate.  Groenewegen et al. (2009) 
give two sets of data, depending on the assumed dust composition. We have adopted the best overall 
fit, which corresponds to the first entry for each object in Table~4 of Groenewegen et al. 
(2009).

The sample of Li-poor stars include then 178 stars in the Gullieuszik sample, 100 stars in the 
C-rich sample by Groenewegen, and 34 stars in their O-rich sample, so that the total sample of Li-poor 
stars has 312 stars. 
\bigskip
\noindent
{\bf 3. The Method}
\medskip\noindent
3.1 Empirical Correlations of Li-rich Stars
\medskip
Maciel \& Costa (2015) considered a large sample of Li-rich giant stars
and investigated the existence of possible correlations involving the Li abundance and
other stellar parameters. In the following we will consider two of these correlations,
namely the correlations of the Li abundance with the stellar radius and luminosity.

\bigskip
\centerline{\psfig{figure=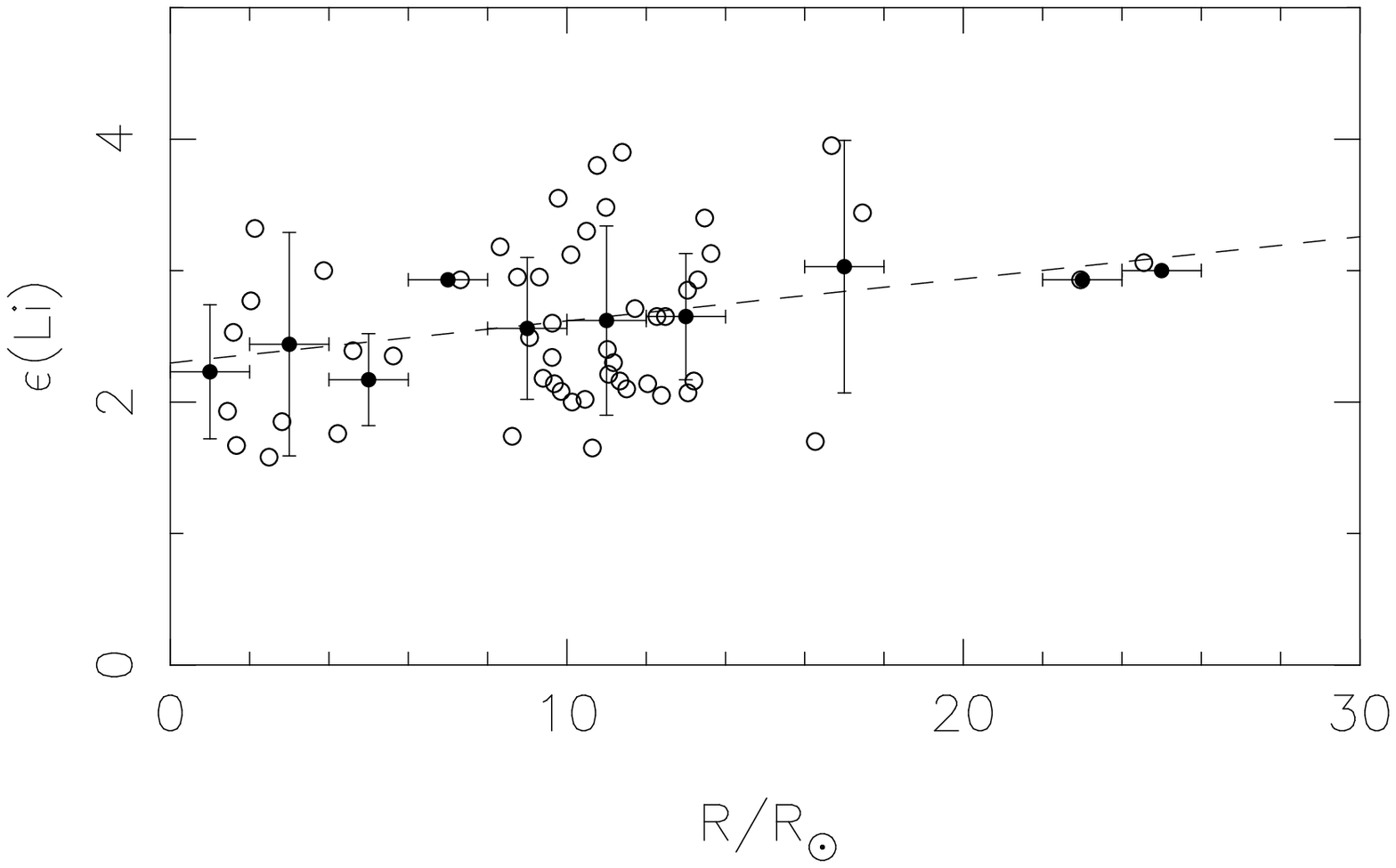,width=8.5 cm,angle=0} }
\medskip
\vbox
{\baselineskip = 8 pt \noindent\sevenbf Figure 1  \sevenrm - Correlation between the Li abundance 
and the stellar radius for Li-rich giant stars. The empty circles show the data used by Maciel 
and Costa (2015), and the black dots are average abundances in 2 solar radius bins. The dashed line
shows the linear fit given by equation~1.}
\bigskip
In our previous work, a relation was obtained in the form $\epsilon{\rm (Li)} = f(R)$,
as shown in Figure~1. In this figure, the empty circles are the Li-rich stars,
the filled dots are average abundances taken in $2\,R_\odot$ bins and the dashed line
shows a linear fit to the data, which is the simplest correlation, and can be considered
as a first approximation. The corresponding equation can be written as

$$\epsilon({\rm Li}) = a + b \ R/R_\odot  \eqno(1)$$

\noindent 
with $a = 2.30 \pm 0.11$, $b = 0.03 \pm 0.01$, and correlation coefficient 
$r = 0.81 \pm 0.20$. The correlation is assumed to  be valid in the interval 
$0 \leq \log R/R_\odot \leq 30 $ and $1.5 \leq \epsilon{\rm (Li)} \leq 4.0 $.

From Figure~1, the Li abundance clearly seems to increase with the stellar
radius, at least within a range that encompasses most the Li-rich stars in the
sample. Explaining the origin of this correlation is a complex procedure, since
the Li enrichment mechanism is not well known, and different models have
been proposed for Li-rich giants near the luminosity bump, clump giants, and
stars on the AGB branch. Some possibilities include incomplete Li dilution,
hot bottom burning, cool bottom burning, among others. We have not addressed this aspect 
in detail, since our main goal here is to compare the mass loss rates of the Li-rich stars with
the majority of Li-poor objects. However, some hints can be made, considering
some recent models for the Li-enrichment in RGB/AGB stars. In particular,
the recent work by Casey et al. (2016) based on 20 Li-rich giant stars from the 
Gaia-ESO survey explains the Li-enrichment process as a natural consequence of
the engulfing of Jupiter-like planets, followed by deep mixing in the stellar envelope
to produce additional Li. This is not proven yet, but in this case the Li-rich objects would 
favour larger stars, which present a larger cross section to absorb the planets, therefore
leading to a $\epsilon(R)$ correlation such as the one shown in Figure~1. 
A similar discussion along these lines was recently presented by Delgado Mena
et al. (2016a,b). Also, recently Kirby et al. (2016) 
proposed a scenario for Li enrichment involving mass transfer from Li-enriched companions 
to RGB stars, a procedure that will also favour larger stars, with an enhanced probability 
of absorbing the companions. As a conclusion, the observed relation is expected, at least 
on the basis of average values, as considered here.

\bigskip
\centerline{\psfig{figure=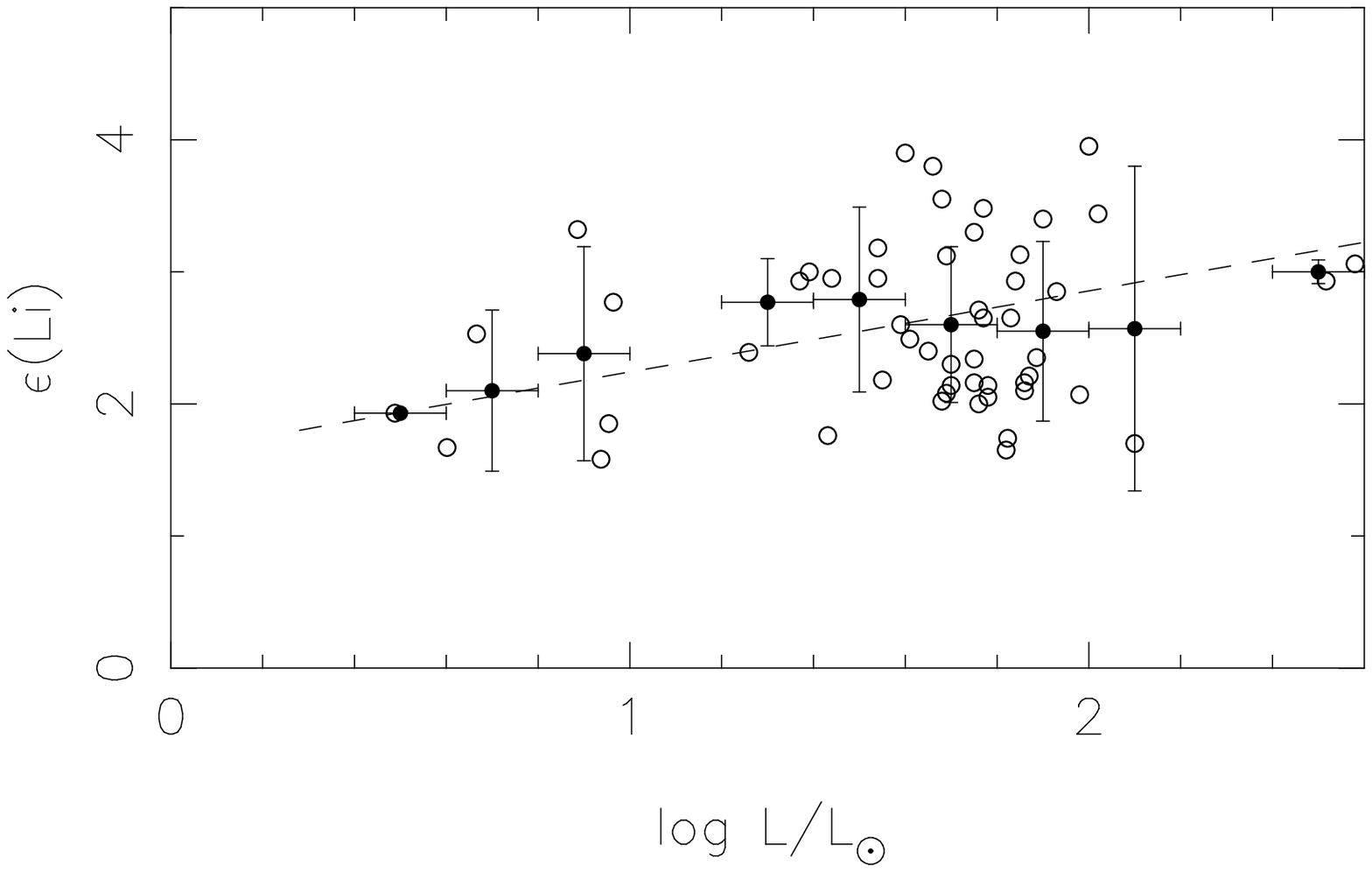,width=8.5 cm,angle=0} }
\medskip
\vbox
{\baselineskip = 8 pt \noindent\sevenbf Figure 2  \sevenrm - Correlation between the Li abundance 
and the stellar luminosity for Li-rich giant stars. The empty circles show the data used by Maciel 
and Costa (2015), and the black dots are average abundances in 0.2 dex luminosity bins. The 
dashed line shows the linear fit given by equation~2.}
\bigskip\medskip

Since we have $L \propto R^2 \,  T_{eff}^4$, the stellar luminosity is approximately proportional
to the radius squared, as the effective temperature does not vary much in the considered sample. 
Therefore, one would then expect a correlation of the stellar luminosity and the Li abundance, 
which we have indeed obtained, as shown in Figure~2, which includes the data in the range 
$0 < \log L/L_\odot < 2.6$ as well as the averages in 9 luminosity bins. Analogously, for the Li 
abundances we have $1.5 \leq \epsilon{\rm (Li)} \leq 4$, but for stars near the lower limit very low
luminosities are obtained, which are outside the range where calculations are
possible, so that we adopt instead the range $1.8 \leq \epsilon{\rm (Li)} \leq 4.0$.
The Li-rich phase is probably a short one in the life of a cool  giant star, and our
sample may include objects in different stages of Li enrichment, which can be seen from
the  scatter in Figure~2. However, an increase of the abundances with the luminosity
is apparent, especially near the upper envelope, so that we feel safe in adopting an
average relation between these quantities.

For practical purposes we will consider the inverse relation as $\log L/L_\odot = 
f[\epsilon{\rm (Li)}]$, and the best correlation obtained for the average data is shown as a 
dashed line in Figure~2, which can  be written as

$$\log L/L_\odot = c + d \ \epsilon{\rm (Li)} \eqno(2) $$

\medskip\noindent
with $c = -2.65 \pm 1.07$, $d = 1.63 \pm 0.42$, and correlation coefficient $r = 0.83 
\pm 0.43$. This equation is strictly valid in the luminosity interval $0 \leq \log L/L_\odot \leq 3.0$,
but we have  found that for the few objects with higher luminosities the results are essentially the 
same, so that a more flexible form of this range  can be written as $0 \leq \log L/L_\odot \leq 5.0$. 
In fact, independent estimates of the luminosity of the Li-rich stars considered here are consistent 
with the adopted range. For example, from the bolometric magnitudes of the stars in the sample by 
Brown et al. (1989) we have $1.1 \leq \log L/L_\odot \leq 2.7$.  Also, from the data by 
Mallik (1999), we get $0 \leq  \log L/L_\odot \leq 3.0$.
\vfill\eject
\noindent
3.2 Determination of the Mass Loss Rates
\medskip
We have developed several methods to derive the mass loss rate of Li-rich AGB/RGB stars, based 
essentially on the Li-abundance and on some correlations as discussed previously by Maciel and Costa 
(2015). In the following we will present a method based on a modified form of the Reimers 
formula, and apply it to the Li-rich sample discussed in section 2. The method can be summarized as 
follows: we adopt the relation between the stellar luminosity and the Li abundance of Li-rich 
stars, as given by equation~2, so that the luminosity can be estimated from the Li abundance. The 
stellar mass is then obtained from the luminosity and effective temperature using recent evolutionary 
tracks. The stellar radius is determined from the stellar gravity, so that the mass loss rate can be 
calculated using an adequate calibration of the Reimers formula involving both Li-rich and Li-poor 
stars in the AGB/RGB branches. Therefore, the mass loss rate depends essentially on (i) the Li 
abundance $\epsilon$(Li), (ii) the effective temperature $T_{eff}$, and (iii) the stellar gravity~$g$.

As a first step, we determine the stellar luminosity using the relation between the luminosity 
and the Li abundance, as given by equation~2. It should be noted that using the relation
between the Li abundance and the stellar radius given by equation~1 we obtain essentially the
same results.  Having the luminosity and the effective temperature, the mass can be estimated using 
recent evolutionary tracks for giant stars. We have adopted the tracks by Bertelli et al. 
(2008), see also Kumar et al. (2011). The tracks can be applied to solar 
metallicity stars with masses in the interval $1.0 < M/M_\odot < 3.0$, and effective temperatures 
in the range $3800 < T_{eff} ({\rm K}) < 5600$. The curves have been approximated by polynomials 
of order 3--6, and are shown in Figure~3. Since the effective temperature and luminosities are 
known and the tracks are reasonably detached, the determination of the stellar mass is a 
straighforward procedure. 

As a further illustration of the relation between the stellar radius and the Li abundance, we have 
included in Figure~3 the average binned data from Figure~1, shown as black dots, where the luminosity 
was estimated from equation~2. The size of the dots is approximately proportional to the Li abundance, 
and the dotted lines show the location of stars with radius $R = 5, \ 10, \ 15, \ {\rm and} \ 20\, 
R_\odot$. It can be seen that the stars with larger Li abundances,  $\epsilon{\rm (Li)} \geq 3.0$, 
are closer to the location of the larger objects.

From the stellar mass and gravity, the stellar radius can be simply calculated as $R^2 = G \ M / g$, 
and the mass loss rate can be estimated by the Reimers formula 

$${dM\over dt} = 4 \times 10^{-13} \ \eta \ {(L/L_\odot)\ (R/R_\odot) \over(M/M_\odot)}
 \eqno(3) $$

\smallskip\noindent
(see for example Lamers and Cassinelli 1999). The parameter $\eta$ was originally 
considered as $\eta \leq 3$, but this calibration is probably not valid for the AGB/RGB stars 
considered here, so that we will consider it as a parameter to be determined. 

\bigskip
\centerline{\psfig{figure=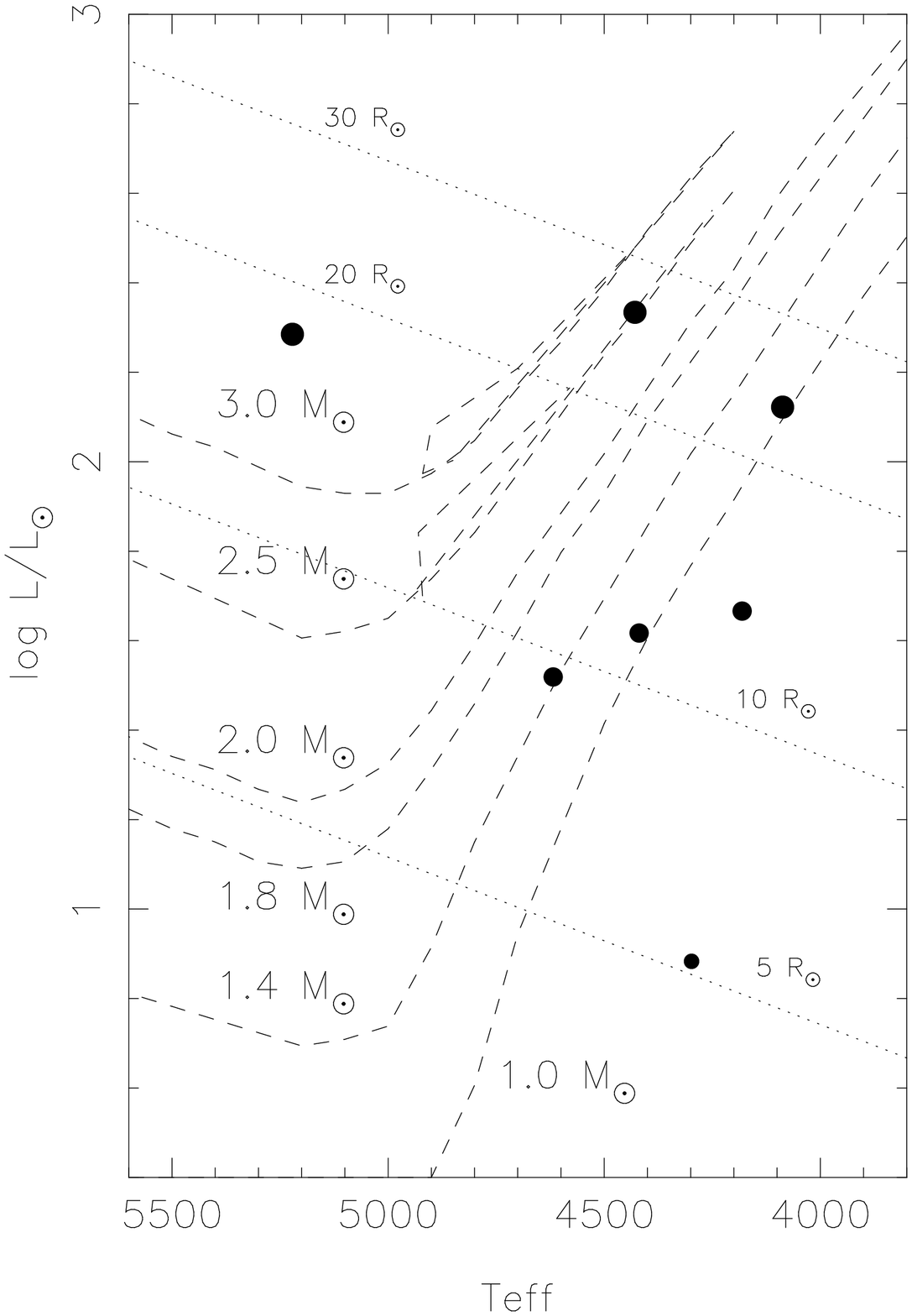,width=7.5 cm,angle=0} }
\medskip
\vbox
{\baselineskip = 8 pt \noindent\sevenbf Figure 3  \sevenrm - Adopted evolutionary tracks from Bertelli 
et al. (2008) for stars in the mass range of 1 to 3 solar masses (dashed lines). The black dots 
show the approximate position on the HR diagram of the binned averages of Figure~1, and the size of 
the dots is proportional to the Li abundance. The dotted lines show the location of stars with 
radii given by  5,  10, 15,  and 20 solar radii.}
\bigskip

\bigskip
\noindent
{\bf 4. Results and Discussion}
\medskip
The adopted value of the parameter $\eta$ was obtained in the  following way: As a first 
approximation, we adopt a linear relation between the mass loss rate $\log dM/dt$ 
and the luminosity $\log L/L_\odot$ for Li-rich stars of the form  $\log dM/dt = 
e + f \, \log L/L_\odot$. The slope of this relation is then obtained using the Li-rich stars, and 
does not depend on the value of $\eta$.  We obtain the following results:  $f = 0.95 \pm 0.04$, 
with $n = 104$ and correlation coefficient $r = 0.91 \pm 0.42$.

Once the slope is fixed, we consider the Li-poor stars for which the mass loss rates and luminosities 
have been independently derived, which include the objects by Gullieuszik et al. 
(2012) and Groenewegen et al. (2009), with a total of 312 
objects. We then use the total sample of 416 objects to determine the best value of the intercept $e$ 
for the $\log dM/dt \times \log L/L_\odot$ correlation, that is, the intercept that corresponds to 
the slope derived for the Li-rich stars.  This  of course determines the parameter $\eta$, since for 
a given star the Reimers formula states that $dM/dt \propto \eta$. We have then for the whole sample 
of 416 stars the results: $\eta =  12.3$,  $e = -10.34 \pm 0.16$, $f = 0.95 \pm 0.05$  with 
correlation coefficient $r = 0.72 \pm 1.03$. 

The effectively used equation can be written as

$${dM \over dt} = 1.74 \times 10^{-12} \ {10^{1.629\,\epsilon({\rm Li})}
    \over (g \, M)^{1/2}} \eqno(4)$$

\noindent
The mass loss rate $dM/dt$ is in  $M_\odot$/year, the Li abundance $\epsilon$(Li) in dex, the gravity 
$g$ in cm/s$^2$, and the stellar mass in solar masses. This equation is of the form 
$dM/dt = f[T_{eff}, g, \epsilon({\rm Li})]$, since the mass depends on the effective temperature and 
on the luminosity, which is determined by the Li abundance. 

\bigskip
\centerline{\psfig{figure=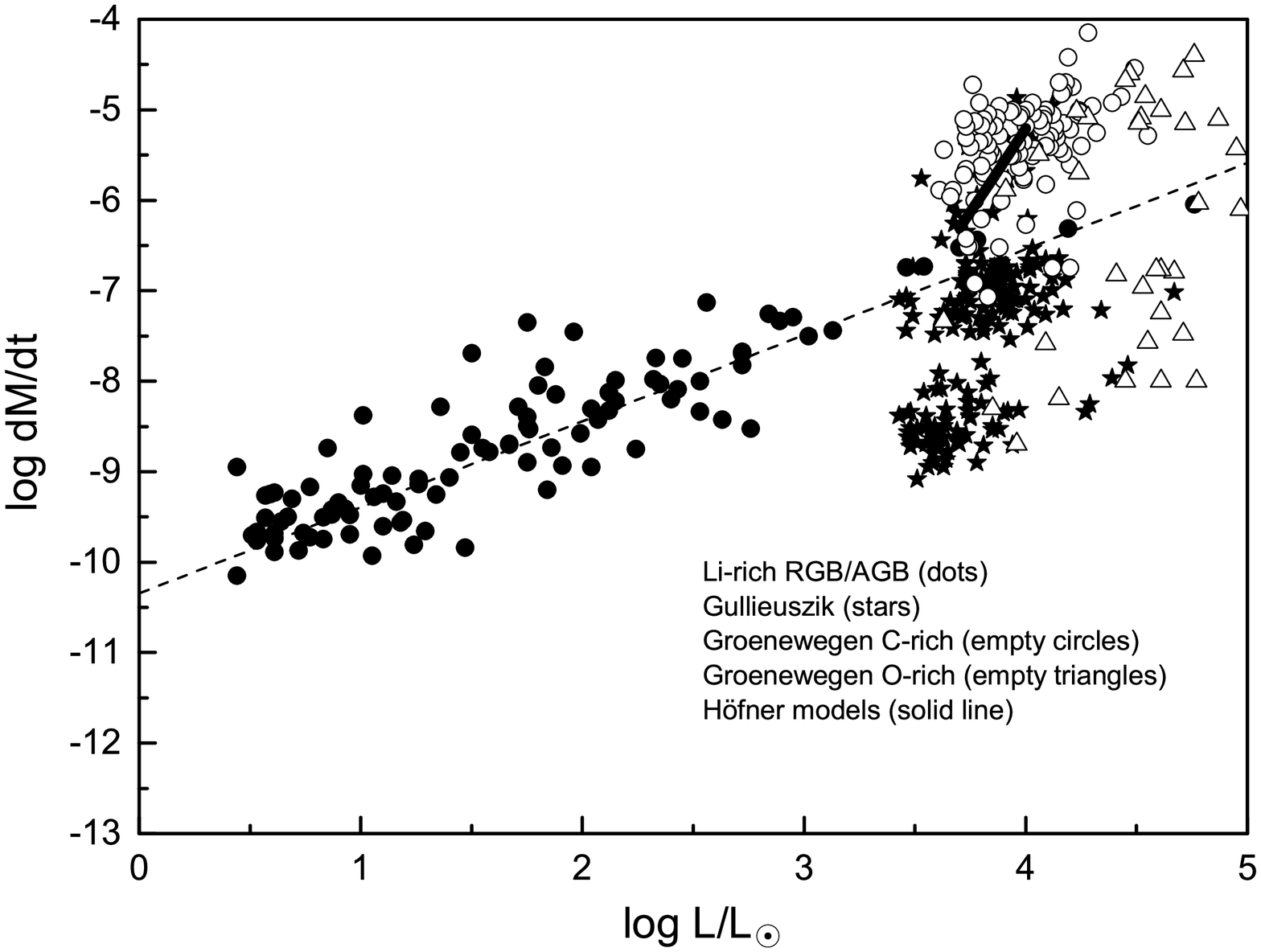,width=12.0 cm,angle=0} }
\medskip
\vbox
{\baselineskip = 8 pt \noindent\sevenbf Figure 4  \sevenrm - Luminosities and mass loss rates 
(solar masses per year) for Li-rich stars (dots) and Li-poor stars (Gullieuszik, stars, Groenwegen 
C-rich, empty circles, Groenewegen O-rich, empty triangles, H\"ofner models, solid thick line.) 
The dashed line shows the derived linear relation.}
\bigskip

Figure~4 shows the $\log dM/dt \times \log L/L_\odot$ plot including Li-rich 
stars (dots); Li-poor objects (Gullieuszik, stars; Groenewegen C-rich, empty circles; Groenewegen 
O-rich, empty triangles,). 

We can make a rough estimate of the uncertainties involved in the determination of the 
mass loss rate by considering the typical uncertainties in the stellar properties
adopted in this work. The uncertainty in the Li abundance $\epsilon$(Li) is typically of 
0.20 dex, according to the original sources cited in Section~2. The effective temperature 
$T_{eff}$ is known to better than 100 K for most objects, and the gravity $\log g$ has a typical 
uncertainty of 0.20 dex. From the adopted correlations involving the stellar radius
$R$ and luminosity $\log L/L_\odot$, we would then expect average uncertainties of about 
$1\,R_\odot$ and 0.20 dex, respectively. This would translate into an uncertainty of
about $0.5\, M_\odot$ for the solar mass, leading to a final uncertainty of about $0.50$ dex
for the mass loss rate $\log dM/dt$, which can also be estimated directly from figure~4. This 
is comparable with the uncertainties in the mass loss rates determined for Li-poor stars by 
Gullieuszik et al. (2012), which show an average dispersion of about 0.5 dex 
for $\log dM/dt$, corresponding roughly to a factor 2 for a typical mass loss rate of 
$dM/dt \sim 10^{-6}\, M_\odot/$year. Groenewegen et al. (2009) quote
a slightly smaller uncertainty of 0.43 dex in $\log dM/dt$ for AGB stars and red supergiants
in the Magellanic Clouds.  It should be stressed, however, that our main point here is not 
the determination of the absolute value of the mass loss rate of Li-rich stars, but to compare
their mass loss rates with those of most Li-poor giants.

H\"ofner \& Andersen (2007) suggested a mass loss mechanism for M-type AGB stars based 
on the formation of carbon and silicate grains due to non-equilibrium effects. It is interesting to 
compare the results of these models with the present results, as shown in Figure~4. The model results 
can be approximately represented by average values of the luminosity and mass loss rates, which are 
included in the figure as a solid line. They are well placed in the upper right part of the 
diagram, where the O-rich giants are located, as expected. It can be seen that the agreement is fair, 
taking into account that the models apply to more luminous stars, having more intense mass loss rates
than the Li-rich objects.

Li-enrichment has been associated with an enhanced mass loss ejection, as discussed by de La Reza et 
al. (1996, 1997). Monaco et al. (2011) comment 
that some Li-rich giants show evidences of mass loss and chromospheric activity, as discussed by  
Balachandran et al. (2000) and Drake et al. (2002). However, 
Fekel \& Watson (1998) and Jasniewicz et al. (1999) suggested that no 
important mass loss phenomena are associated with these stars. By a comparison of K - [12 $\mu$m] 
colours of the 3 Li-rich stars with corresponding data for high mass loss Miras, Lebzelter et al. 
(2012) suggested that the Li-rich objects do not have enhanced dust mass loss. An 
enhanced gas mass loss has also been ruled out by the lack of asymmetries in the Ha profile, so that 
the conclusions by Fekel \& Watson (1998) and Jasniewicz et al. (1999) 
are supported. In fact, as pointed out by Mallik (1999) and  Luck (1977), 
a large amount of mass loss would remove the stellar outer layers where most Li atoms are located, 
so that strong mass loss rates are probably not associated with Li excess in AGB/RGB stars. The 
results shown in Figure~4 also confirm that the Li enrichment process does not seem to 
be associated with particularly strong mass loss rates, except for very high luminosity stars, which 
are a small fraction of the known Li-rich AGB/RGB stars.

\bigskip
\noindent
{\bf Acknowledgements}
\bigskip
\noindent
This work was partially supported by FAPESP and CNPq.

\bigskip
\noindent
{\bf References}
\bigskip
\par\noindent

\par\noindent
Balachandran, S. C., Fekel, F. C., Henry, G. W., Uitenbroek, H. 2000, ApJ 542, 978

\par\noindent
Bertelli, G., Girardi, L., Marigo, P., Nasi, E. 2008, A\&A 484, 815

\par\noindent
Brown, J. A., Sneden, C., Lambert, D. L., Dutchover, E. 1989, ApJ Suppl. 71, 293

\par\noindent
Cameron, A. G. W., Fowler, W. A. 1971, ApJ 164, 111

\par\noindent
Casey, A. R., Ruchti, G., Masseroni, T., et al. 2016, MNRAS 461, 3336

\par\noindent
Charbonnel, C., Balachandran, S. C. 2000, A\&A 359, 563

\par\noindent
de La Reza, R., Drake, N. A., da Silva, L. 1996, ApJ 456, L115

\par\noindent
de La Reza, R., Drake, N. A., da Silva, L., et al. 1997, ApJ 482, L77

\par\noindent
Delgado Mena, E., Tsantaki, et al. 2016a, Cool Stars 19, Ed. G. A. Feiden

\par\noindent
Delgado Mena, E., Tsantaki, M., et al. 2016b, A\&A 587, A66

\par\noindent
Drake, N. A., de la Reza, R., da Silva, L., Lambert, D. L. 2002, AJ 123, 2703

\par\noindent
Fekel, F. C., Watson, L. C. 1998, AJ 116, 2466

\par\noindent
Gonzalez, O. A., Zoccali, M., Monaco, L., et al. 2009, A\&A 508, 289

\par\noindent
Groenewegen, M. A. T., Sloan, G. C., Soszynski, I., Peterson, E. A. 2009, A\&A  506, 1277

\par\noindent
Groenewegen, M. A. T., Wood, P. R., Sloan, G. C., et al. 2007, MNRAS 376, 313

\par\noindent
Gullieuszik, M., Groenewegen, M. A. T., Cioni, M. R. L., et al.  2012, A\&A  537, A105

\par\noindent
H\"ofner, S., Andersen, C. A. 2007, A\&A  465, L39

\par\noindent
Jasniewicz, G., Parthasarathy, M., de Laverny, P., Th\'evenin, F. 1999, A\&A  342, 831

\par\noindent
Kirby, E. N., Guhathakurta, P., Zhang, A. J., et al.  2016, ApJ 819, 135

\par\noindent
K\"ov\'ari, Zs., Korhonen, H., Strassmeier, K. G., et al. 2013, A\&A  551, A2

\par\noindent
Kumar, Y. B., Reddy, B. E., Lambert, D. L. 2011, ApJ 70, L12

\par\noindent
Lamers, H. J. G. L., Cassinelli, J. 1999, Introduction to stellar winds, Cambridge 

\par\noindent
Lebzelter, T., Uttenthaler, S., Busso, M., Schultheis, M., Aringer, B. 2012, A\&A  538, 36

\par\noindent
Luck, R. E. 1977, ApJ 218, 752

\par\noindent
Lyubimkov, L. S., Lambert, D. L., Kaminsky, B. M., et al.  2012, MNRAS 427, 11

\par\noindent
Maciel, W. J., Costa, R. D. D. 2012, Mem. S. A. It. S. 22, 103

\par\noindent
Maciel, W. J., Costa, R.D.D. 2015, Why galaxies care about AGB stars, ASP CS 497, 313

\par\noindent
Mallik, S. V. 1999, A\&A 352, 495

\par\noindent
Martell, S. L., Shetrone, M. D. 2013, MNRAS 430, 611

\par\noindent
Monaco, L., Villanova, S., Moni Bidin, C., et al.  2011, A\&A 529, A90

\par\noindent
Monaco, L., Boffin, H. M. J., Bonifacio, P., et al.  2014, A\&A 564, L6

\par\noindent
Vassiliadis, E., Wood, P. R. 1993, ApJ 413, 641

\par\noindent
Wood, P. R., Bessell, M. S., Fox, H. W. 1983, ApJ 272, 99 33

\bye